\documentclass[checkin,aps,pra]{revtex4}

\usepackage{makeidx}
\usepackage{listings}
\newcommand{\ba}{\begin{eqnarray}}
\newcommand{\ea}{\end{eqnarray}}

\usepackage{graphicx,color}

\newcommand{\ce}{$\rm C_{60}$}

\newcommand{\be}{\begin{equation}}
\newcommand{\ee}{\end{equation}}
\newcommand{\la}{\langle}

\newcommand{\ra}{\rangle}
\newcommand{\et}{{\it et al. }}

\newcommand{\clr}{}

\begin{document}




\title{Laser-induced forces on atoms during ultrafast demagnetization}





\author{G. P. Zhang$^*$} \affiliation{Department
  of Physics, Indiana State University, Terre Haute, Indiana 47809,
  USA}


\author{Y. H. Bai} \affiliation{Office of Information
  Technology, Indiana State University, Terre Haute, Indiana 47809,
  USA}

\date{\today}

\begin{abstract}
  { Laser-induced femtosecond demagnetization has attracted a broad
    attention as a possible candidate for information storage
    technology. However, whether or not lattice vibration directly
    participates in demagnetization has been highly controversial over a
    decade. A recent electron diffraction experiment attributed the
    demagnetization to the polarized phonon effect, but a similar
    x-ray diffraction experiment attributed it to the Einstein-de Haas
    effect. Common to both experiments is that neither the angular
    momentum of the lattice nor the rotation of the sample was
    directly probed. Here, we report our first first-principles
    calculation of forces on atoms induced by an ultrafast laser
    during ultrafast demagnetization. We employ two complementary
    methods: (i) the frozen lattice with electronic excitation and
    (ii) frozen excitation but moving the lattice.  We find that the
    forces on atoms start at -50 fs and peak around 30 fs. The
    magnitude of the force is far smaller than the empirical
    estimates. Within the limit of our theory, our results suggest
    that the polarized phonon effect and the Einstein-de Haas effect
    are unlikely to be the main course of demagnetization.  We expect
    that our finding has a profound impact on the future direction of
    laser-induced dynamics in magnetic and quantum materials. }
\end{abstract}


 \maketitle

 \section{Introduction}

In 1996, Beaurepaire and his coworkers \cite{eric} reported that a
60-fs laser pulse can quench the magnetization of a nickel thin film
more than 40\% within 1 ps. This pioneering discovery has forever
changed the landscape of spin manipulation, where a laser field,
instead of a magnetic field, is employed. The accelerated
demagnetization, found in many magnetic materials
\cite{hohlfeld1997,aeschlimann1997,ju1998,roth2012,mann2012,sultan2012,chen2019},
has attracted enormous attentions across several decades around the
world\cite{ourreview,rasingreview}, for a historical account of the
discovery of femtomagnetism, see a recent book \cite{ourbook}.  This
unprecedented ultrashort time scale involves a large group of
interactions among electrons, spins, orbitals and phonons, either
directly or indirectly, which significantly complicates femtomagnetism
\cite{ourreview}.  The first theory \cite{prl00} emphasized a crucial
interplay between the laser field and the spin-orbit coupling, so the
spin symmetry can be broken and the spin moment can be
changed. Subsequent mechanisms invoke the phonon-based Elliot-Yafet
mechanism \cite{koopmans2010,born2021} with the spin flipping through
the spin-orbit coupling \cite{prl00}, spin transport with majority
spin moving out of the excited region \cite{battiato2012}, bands
mirroring effect \cite{eich2017} with the spin majority and minority
switching their roles, spin-wave excitation \cite{jap19} across many
lattice sites to reduce the spin moment for the entire sample, spin
disordering \cite{chen2019w} within a big cell, ultrafast electron
correlations and memory effects \cite{prl20} through the dynamic
exchange kernel, optical inter-site spin transfer \cite{willems2020},
and electron-phonon scattering \cite{baral2014}. The fact that so many
mechanisms were proposed demonstrates the complexity of
femtomagnetism. To this end, no single mechanism can explain all the
experimental observations; and probably it may never be possible,
given the fact that diverse materials show different dynamics. For
instance, Fe and Gd have quite different demagnetization processes
\cite{sultan2012}.

Another line of thought is formulated around the spin angular momentum
transfer \cite{hennecke2019,dewhurst2021b,zahn2021}, without
considering the fact that the total angular momentum operator is not a
conserved quantity in solids \cite{prb08}.  Once this total angular
momentum conservation is assumed, then it is easy to see that if the
spin angular momentum of the electron system cannot transfer to the
orbital angular momentum of the electron \cite{jpcc21}, it must go to
the lattice \cite{stamm2010}. Although this mechanism cannot exclude
other mechanisms such as spin quenching due to the spin misalignment,
it underlines two recent experimental investigations
\cite{dornes2019,tauchert2022}. However, to this end, no experiment
ever {\it directly} measured the rotation of samples or the lattice
angular momentum as one should if this is due to the Einstein-de Haas
effect.  Dornes \et \cite{dornes2019} employed the time-resolved X-ray
diffraction as the probe and 40-fs laser pulse of 800 nm as the pump
and measured the transverse strain wave, from which they inferred
through modeling that 80\% spin angular momentum in the Fe film went
to the lattice within 200 fs.  Tauchert \et \cite{tauchert2022}
carried out the electron diffraction experiment and found the Bragg
spot position from their Ni film did not change with time within their
experimental uncertainty, so there was no transient strain or lattice
expansion, contradicting the finding by Dornes \et \cite{dornes2019}.
It is difficult to argue that the different conclusions reached from
these two experiments are due to the technique difference, as electron
diffraction should induce more strain than X-ray diffraction.
Instead, Tauchert \et \cite{tauchert2022} found an anisotropic
diffraction pattern that follows the initial magnetization of their
sample, which they attributed to the polarized phonon involvement, but
phonons are rarely magnetically active \cite{baydin2022} and the
long-lasting associated orbital moment of the electron observed in He
atoms \cite{watzel2022} cast doubts on this interpretation.  The key
to resolve these difficult issues is to find the force on the
atoms. If the force is centripetal, then the postulate of the
polarized phonon is valid.

Theory can help in this regard by tracing back to the initial stage of
laser excitation, where electrons are newly excited out of the Fermi
sea and the force starts to appear on atoms. In this paper, we develop
a scheme that combines the ground state density functional theory and
the time-dependent Liouville formalism in a supercell. This allows us
to compute the force on the atoms during the laser-induced ultrafast
demagnetization.  We employ one free-standing Ni(001) monolayer as our
first example. We find that the force on the Ni atom starts around -50
fs, followed by a rapid oscillation due to the strong charge
fluctuation. The force ceases to change once the electronic excitation
configuration is formed. Quantitatively, we find that under moderately
strong laser pulse excitation, the amplitude of force is 0.01
mRy/a.u., which is far below the estimate based on the Tauchert's
parameter \cite{tauchert2022}.  Although the force does change
directions during the laser excitation, we do not detect evidence of
the centripetal force as required by the polarized phonon. Our results
in fcc Ni further show that the directions of forces on all four Ni atoms
do not change. Therefore, our present theoretical results, within the
limit of our current theory, do not support the picture of the
polarized phonons. This finding is expected to motivate further
experimental and theoretical investigations in the future.

The rest of the paper is arranged as follows.  In Sec. II, we present
our theoretical algorithm, which combines the ground state calculation
and the time-dependent Liouville equation. Section III is devoted to
(i) the results in the Ni(001) monolayer, (ii) the effect of core
states and the spin polarization on the force, and (iii) the results
in fcc Ni. We conclude this paper in Sec. IV.

\newcommand{\br}{{\bf r}}

\newcommand{\ik}{i{\bf k}}

\newcommand{\jk}{j{\bf k}}

\newcommand{\lk}{l{\bf k}}

\newcommand{\bk}{{\bf k}}

\section{Theoretical formalism}

\newcommand{\iik}{i,i,{\bf k}}

\begin{figure}
  \includegraphics[angle=0,width=0.9\columnwidth]{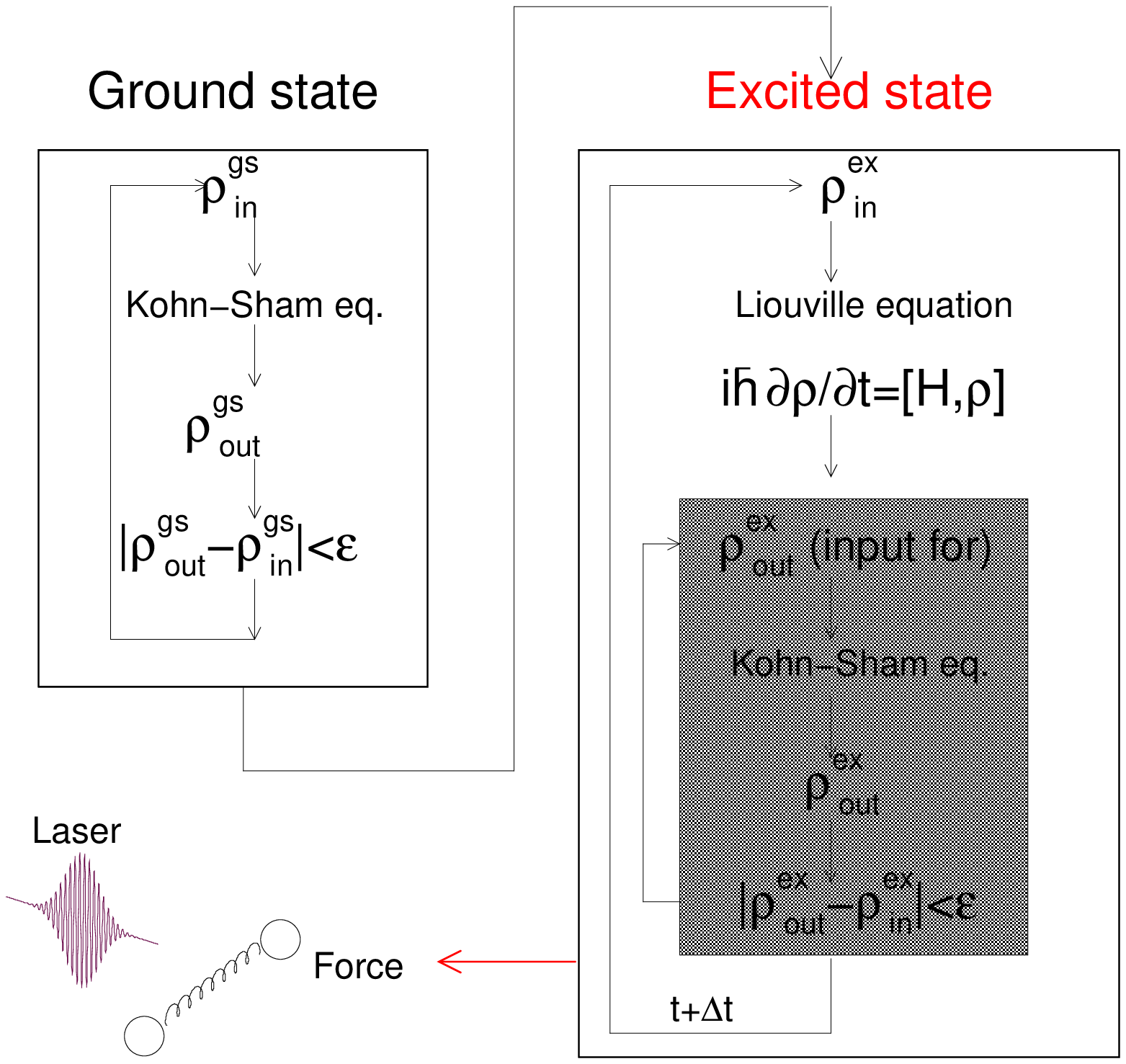}
  \caption{ (Bottom left) Laser-induced force on atoms during the
    ultrafast demagnetization has an enormous impact on subsequent
    lattice dynamics.  We employ a circularly polarized 60-fs laser
    pulse $\sigma_{xy}$.  For a Ni(001) monolayer, a supercell
    $(2\times 2\times 1)$ is used.  (Top left) Schematic of our
    theoretical formalism for our ground state calculation. (Right)
    Time-dependent self-consistent Liouville density functional
    theory.
}
\label{fig0}
  \end{figure}

Laser-driven ultrafast demagnetization starts with the electronic
excitation which tip the balance of force on atoms.  Figure \ref{fig0}
schematically shows a typical laser excitation. A laser pulse excites
the electron first and through the electron-lattice interaction, atoms
experience a net force. So the key to resolve the above controversy is
to directly compute the forces on atoms under laser excitation.  If
the force is centripetal, then the atoms will undergo a circular
motion. However, such a calculation is difficult, if not impossible,
since a solid has highly coupled interactions among electron, spin and
lattice dynamics over several hundred fs in a supercell. This is part
of the reason why so far there has been no theory attempt to compute
forces. We find a method that integrates two complementary algorithms
into a well-defined formalism.

The top left of Fig. \ref{fig0} shows that our density-functional
theoretical calculation starts with the ground state calculation, by
solving the Kohn-Sham equation self-consistently \cite{jpcm16}, \be
\left [-\frac{\hbar^2\nabla^2}{2m_e}+v_{eff}(\br) \right ]
\psi_{\ik}(\br)=E_{\ik} \psi_{\ik} (\br),
\label{ks}
\ee where $ \psi_{\ik}(\br)$ and $E_{\ik}$ are, respectively, the
eigenstate and eigenenergy of band $i$ and ${\bf k}$ point.  $v_{eff}$
is determined by \be v_{eff}({\bf r})=v({\bf r})+\int \frac{n({\bf
    r}')}{|{\bf r}-{\bf r}'|} d{\bf r}'+v_{xc}({\bf r}), \ee where
$v_{xc}({\bf r})$ is the exchange-correlation potential, $v_{xc}({\bf
  r})=\delta E_{xc}[n]/\delta n({\bf r})$. We employ the Wien2k code,
which uses the full-potential augmented planewave basis.  The
spin-orbit coupling is included through the second variational
principle \cite{wien2k}.  We use the generalized gradient
approximation for the exchange-correlation energy functional.



The forces on atom $\alpha$ contain several contributions. The
Hellmann-Feynman (HF) force in the atomic units is \cite{yu1991} \be
{\bf F}^\alpha =z_\alpha \frac{d}{d{\tau}_\alpha} \left (
-\sum'_{\beta} \sum_{\bf R}
\frac{z_\beta}{|\tau_\alpha-\tau_\beta+{\bf R}|} + \int
\frac{n(\br)}{|\tau_\alpha-\br|}d\br \right ), \ee where $z_\alpha$ is
the atomic number, $\tau_\alpha$ is its position in a unit cell, and
${\bf R}$ is the lattice vector. The prime over the summation excludes
the same atom. This force is common among all the methods. In LAPW,
the states are separated into core states, semicore state, and valence
states.  The core correction to the HF force is \be {\bf F}_{\rm
  core}^{\alpha}=-\int \rho_c^\alpha (\br) \nabla v_{eff}(\br) d\br,
\label{core}
\ee where $\rho_c^\alpha (\br)$ is the core charge density at atom
$\alpha$. The treatment of valence states is more complicated due to
the incomplete basis-function set (IBS) that is associated with the
LAPW basis function, \be {\bf F}_{\rm IBS}=-\sum_i\rho_i\left ( \la
\psi_i'|(H-\epsilon_i)|\psi_i\ra +\la \psi_i|(H-\epsilon_i)|\psi_i'\ra
+{\bf D}_i \right ), \ee where $\rho_i$ is the occupation number of
valence state $i$, $\epsilon_i$ is the eigenvalue, $H$ is the
Kohn-Sham Hamiltonian, $\psi_i$ is the eigenfunction, and $\psi_i'$
denotes its derivative with respect to the atomic position
$\tau_\alpha$. ${\bf D}_i$ is the kinetic energy contribution due to
the change in both the MT sphere and the interstitial regions
\cite{yu1991,madsen2001}.  The total force is the sum of the HF force,
the core and IBS corrections.  Wien2k has a limitation that its force
has no contribution from the spin-orbit coupling (SOC) even if we
include SOC. Since SOC in general is a weak contributor energetically,
we do not expect that missing force from SOC has a qualitative effect
on our force.

Upon laser excitation, electrons are first excited out of the Fermi
sea. We employ the time-dependent Liouville equation \cite{np09,jpcm16}
\be i\hbar
\frac{\partial \rho}{\partial t} = [H_0+H_I,\rho],
\label{liou}
\ee where  $H_0$ is the
unperturbed system Hamiltonian and $H_I$ is the interaction between
the laser field and the system \cite{np09}, $H_I=-{{\bf p}\cdot{\bf
    A}(t)}/{m_e}$. The vector potential is ${\bf
  A}(t)=A_0\exp(-t^2/\tau^2)(\cos\omega t \hat{x} + \sin\omega t
\hat{y} )$, where $A_0$ is the amplitude, $\tau$ is the laser pulse
duration, $t$ is the time, $\hat{x}$ and $\hat{y}$ are the unit
vectors of the $x$ and $y$ axes, respectively, and the laser
polarization is within the $xy$ plane.   {\clr We note in passing that
  the fluence of laser field is computed from \cite{ourbook}
  \be
  {\cal F}(\omega,\tau)=\sqrt{\frac{\pi}{2}}n(\omega)c\epsilon_0A_0^2\omega^2\tau,\label{F}
  \ee
where $c$ is the speed of light, $n(\omega)$ is the index of
refraction, and $\tau$ is the pulse duration.}

We only propagate our time step from $t$ to $t+\Delta t$.  The right
figure of Fig. \ref{fig0} outlines the key steps of our
implementation. The excited state density $\rho$ is used to construct
the excited state density and is fed back into Eq. \ref{ks} for an
excited state self-consistent calculation \cite{jpcm16}. The shaded
box highlights this self-consistency. The converged density is looped
back into the Liouville equation (Eq. \ref{liou}) to start a new time
step.

\section{Results and  discussions}


Traditional force calculations in solids often adopt a single $\Gamma$
point \cite{liu2021}, which is apparently unsuitable for metals.  We
propose a different formalism to tackle this issue. We freeze the
lattice in the same fashion as the frozen phonon calculation, and then
directly compute the forces on the atoms during laser
excitation. Regardless of whether angular momentum transfer or
electron-lattice coupling is central to demagnetization, the lattice
must experience a force in order to vibrate.

\subsection{Nickel(001) monolayer}

We take a free-standing nickel(001) monolayer as an example. Since its
primitive cell has no net force on atoms, we adopt a $2\times 2\times
1$ supercell, with two distinctive atoms Ni$_1$ and Ni$_2$ at (0,0,0)
and (1/2,1/2,0), respectively.  We insert a vacuum layer of thickness
33 bohr along the $c$ axis to eliminate the interaction between each
layer. We choose a $k$ mesh of $20\times 20\times 1$ in the crystal
momentum space.  The product of the muffin-tin radius and the
planewave cutoff $R_{\rm MT}K_{\rm max}$ is 7.

\begin{figure}
  \includegraphics[angle=0,width=1\columnwidth]{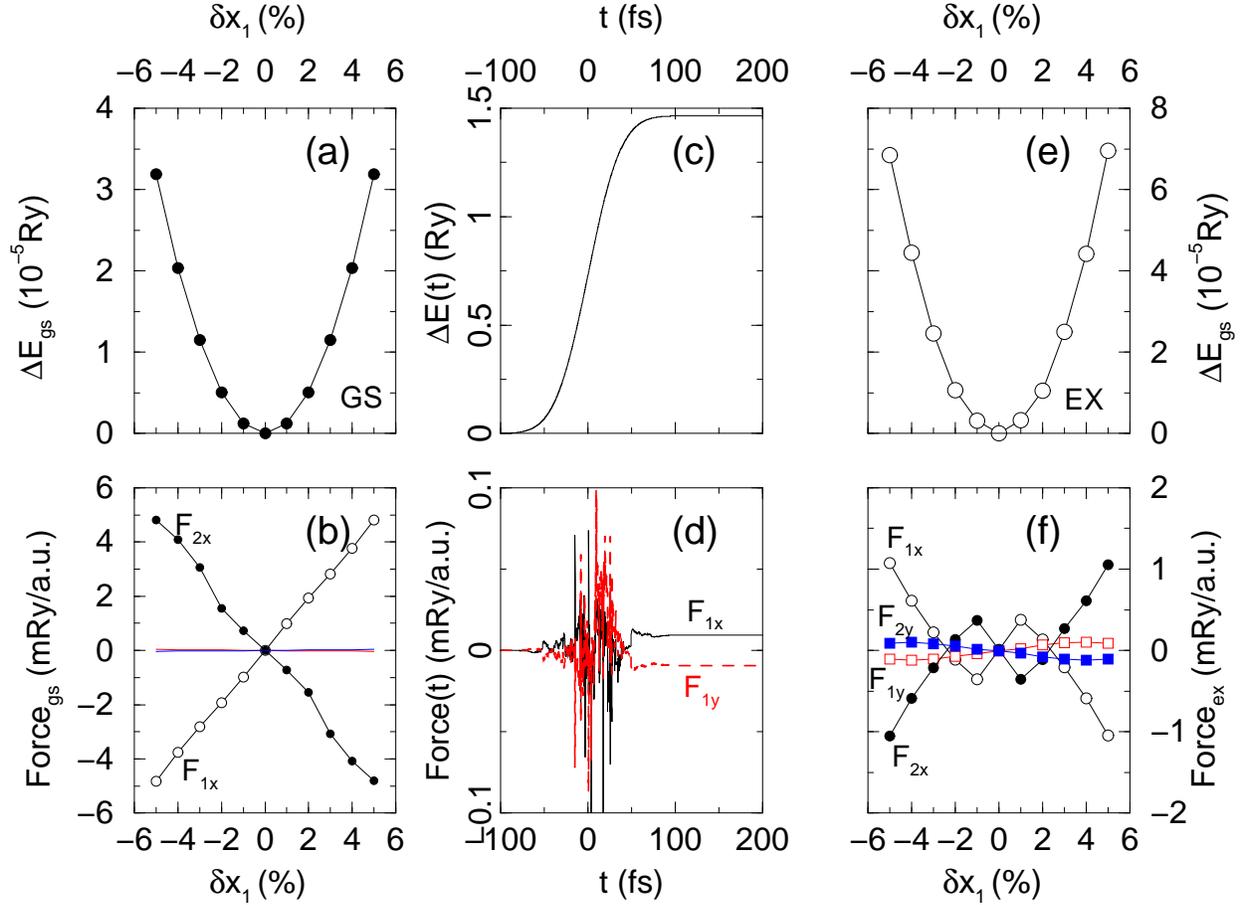}
  \caption{(a) Total energy change $\Delta E_{gs}$ as a function of
    the Ni$_1$ position $\delta x_1$ in the electronic ground state.
    $\delta x_1$ is in the units of lattice constant $a$.  (b) Force
    as a function of $\delta x_1$.  The open circles denote $F_{1x}$
    and the filled circles denote $F_{2x}$. The $y$ components are
    small, close to the zero. (c) Under laser excitation, the total
    energy change $\Delta E(t)$ as a function of time $t$. The laser
    pulse has duration of 60 fs, photon energy of 1.6 eV and the
    vector potential amplitude of 0.03 $\rm Vfs/\AA$. (d) Force on
    Ni$_1$. The solid line represents $F_{1x}$ and the dotted line
    $F_{1y}$. Force on Ni$_2$ is similar, and not shown. (e) In the
    electronic excited states, the total energy change $\Delta E_{ex}$
    as a function of $\delta x_1$. (f) In the electronic excited
    states, the forces on atoms have a different dependence from the
    ground state counterpart.  The unfilled circles denote $F_{1x}$,
    and the filled ones $F_{2x}$. The unfilled and filled boxes
    correspond to $F_{1y}$ and $F_{2y}$, respectively.  }
\label{fig1}
  \end{figure}

{\it Forces in the ground state --} We displace the Ni$_1$ atom along
the $x$ axis by a small amount $\delta x_1$ from -5\% to +5\%, in the
unit of the lattice constant $a$. Figure \ref{fig1}(a) displays the
total energy change $\Delta E$ as a function of $\delta x_1$, where a
typical parabola is observed. Here the energy is referenced to the
minimum energy. Forces on atoms are shown in Fig. \ref{fig1}(b). Since
we move Ni$_1$ along the $x$ axis, the main force is along the $x$
axis, while the forces along the $y$ are tiny (see the lines close to
0). The empty (filled) circles denote the force $F_{1x}$ ($F_{2x}$) on
Ni$_1$ (Ni$_2$).  $F_{1x}$ and $F_{2x}$ have an opposite slope as
expected. The force slope gives an estimate of the angular frequency
of 10.76 THz for Ni, representing an upper bound phonon frequency in
the electronic ground state.

{\it Forces in the excited state --} We employ a circularly polarized
pulse of $\tau=60$ fs and the photon energy is $\hbar\omega=1.6$ eV.
We choose a vector field amplitude $A_0=0.03\rm ~Vfs/\AA$, {\clr which
  corresponds to a fluence of 10.61 $\rm mJ/cm^2$, very typical
  experimentally \cite{eric,jpcm10,dornes2019,tauchert2022}. The
  full-width at half-maximum of the pulse is about 100 fs.}  Figure
\ref{fig1}(c) is the energy change as a function of time. We see the
system absorbs the energy quickly upon laser excitation; the entire
process starts at -100 fs and ends around 100 fs. Our interest is in
the force. Figure \ref{fig1}(d) shows the force on Ni$_1$, and the
force on Ni$_2$ is similar and not shown. $F_{1x}$ gains an
appreciable value around -50 fs, and then it starts a rapid
oscillation. We verify that these rapid oscillations are from the
charge fluctuation during the time propagation of density. The spikes
seen in the force in the self-consistent step appear from one time
step to next. Our time step is 1/32 the laser period. We run four
separate calculations under different conditions, with or without the
core states and with or without spin-polarization rescaling
\cite{jpcm16}, to verify that while details of these rapid
oscillations may differ from one to another, the general pattern is
highly reproducible. Naturally, atoms cannot respond to these rapid
oscillations, and the force after these rapid oscillations is
important. If the polarized phonon concept \cite{tauchert2022} is
valid, we are supposed to see the force direction change
periodically. This direction change is barely observed around 42 fs,
where $F_{1y}$ transitions from a positive force to a negative one,
while $F_{1x}$ does the opposite. However, after 50 fs, both force
components settle down, and there is no evidence of centripetal force
from our data.

{\clr  Tauchert \et \cite{tauchert2022} carried out ultrafast electron
 diffraction to detect the motion of Ni atoms in a nickel single
 crystal layer after optical excitation.  They found no change in the
 Bragg spot broadening and no change in lattice displacement, but the
 intensity ratio between (200) and (020) planes depends on the
 direction of the applied magnetic field. They attributed this
 difference to the polarized phonon. }
They assumed that the 50\% spin angular
 momentum loss is converted to the circular motion of the Ni atom, \be
 \Delta L =M_{\rm Ni} \omega R^2=0.16\hbar, \label{eq3} \ee where
 $\omega$ is angular frequency, $M_{\rm Ni}$ is the atomic mass of Ni,
 and $R$ is the radius of the circular motion of 0.019 $\rm \AA$
 \cite{tauchert2022}. Because the centripetal force on the atom is \be
 F=M_{\rm Ni}\omega^2R, \label{eq4} \ee we combine Eqs. \ref{eq3} and
 \ref{eq4} to rewrite the force as \be F=\omega \Delta L/R. \ee This
 expression has a nice feature where the force is independent of mass,
 thus eliminating the material difference between Dornes
 \cite{dornes2019} and Tauchert \cite{tauchert2022}. The only
 remaining quantity is $\omega$.  Unfortunately, there is no
 information on $\omega$. Due to the translation symmetry in solids, a
 single normal mode never undergoes a circular motion, qualitatively
 different from molecular crystals \cite{prb21}.  Tauchert \et
 \cite{tauchert2022} interpreted the highest phonon frequency of 8 THz
 found in the nickel phonon spectrum as $\omega$, which in general is
 incorrect or questionable.  But using it, we can estimate the force
 from their data to be 10.84 mRy/bohr per atom, a value that we can
 compare with.


Figure \ref{fig1}(d) presents a different picture. Our force is only
0.01 mRy/bohr, three orders of magnitude less than theirs.  10.84
mRy/bohr is also too large for the ground state. If one used the same
slope of the force as Fig. \ref{fig1}(b), 10.84 mRy/bohr would
translate $\delta x_1$ to be 10\%, or 0.3 $\rm \AA$, which is really
too large for nuclear vibrations. In \ce, we find\cite{prb21} the
lattice displacement around 0.05 $\rm \AA$.  Up to now, our lattice is
frozen. What if we release it in the electronic excited state? We
choose an electronic excited configuration around 566 fs, and then
displace Ni$_1$ along the $x$ axis, just as we did for the ground
state in Fig. \ref{fig1}(b). Figure \ref{fig1}(e) shows the total
energy change as a function of $\delta x_1$. We find that in the
electronic excited state, the parabola is steeper. For the same
$\delta x_1$, $\Delta E_{ex}=7\times 10^{-5}$ Ry, larger than $\Delta
E_{gs}=3.2\times 10^{-5}$ Ry. The force changes are shown in
Fig. \ref{fig1}(f). It is clear that the force change is more complex
and no longer has a linear relation with $\delta x_1$. If $\delta x_1$
is smaller than $\pm 1\%$, the dependence is similar to the ground
state (see Fig. \ref{fig1}(b)). But as soon as $\delta x_1$ is larger
than $\pm 2\%$, $F_{1x}$ decreases with $\delta x_1$. $F_{2x}$ shows
an opposite trend. In the ground state, $F_{1y}$ and $F_{2y}$ are very
small, but now in the excited state, they become appreciable. If we
compare the force magnitudes between Figs. \ref{fig1}(d) and
\ref{fig1}(f), we see that the force reaches 1 mRy/bohr.  {\clr This
  shows that our excited state force in Fig. \ref{fig1}(d) is a lower
  limit}. But the force at the maximum is still 10 times smaller than
Tauchert's force.

\subsection{Effect of core states and spin polarization  scaling on the force}

Since there has been no prior study on the laser-induced force on
atoms during demagnetization, we carry out four complementary tests to
iron out some of possible scenarios that Wien2k could provide.

\begin{figure}
  \includegraphics[angle=0,width=1\columnwidth]{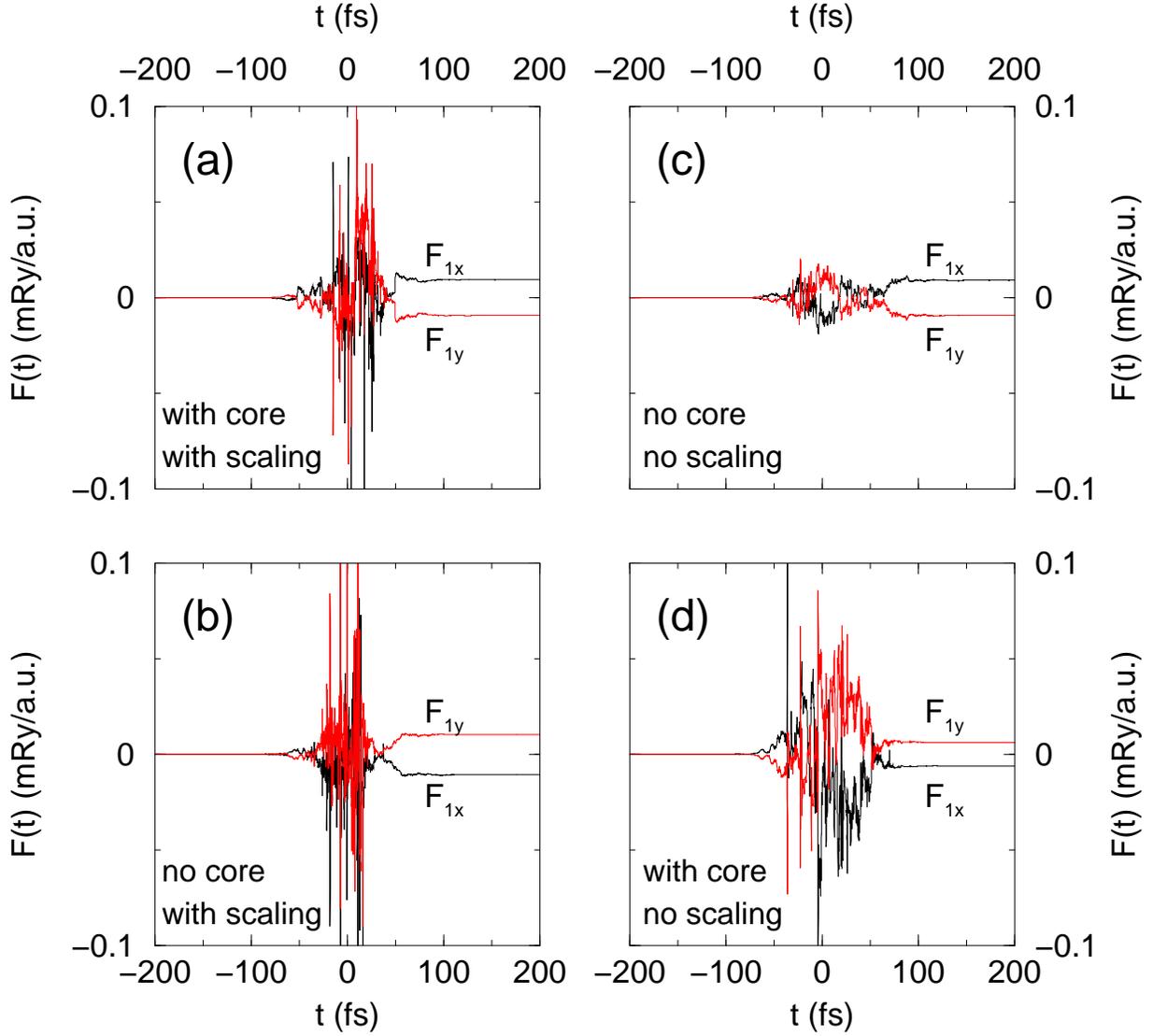}
  \caption{Forces on atoms under four different conditions.  (a)
    Energy window is between [5,100], where the core states are
    included and the spin-polarization rescaling is used. (b) Energy
    window is between [17,120] and does not include the core
    states. The spin rescaling is used. (c) Energy window is between
    [17,120], where the core states are not included and the
    spin-polarization rescaling is not used. (d) Energy window is
    between [5,120], where the core states are included and the
    spin-polarization rescaling is not used. }
\label{fig3}
  \end{figure}

We first examine the effect of core states on our force.  The Ni-L
edge has eight core states ($3s^23p^6$) for each Ni atom, which locate
65 eV below the Fermi level. There are 10 valence electrons
($3d^84s^2$ or $3d^94s^1$).  A single excitation from 1.6-eV photons
rarely affects the core states, but multiple excitations do. Since
core states have a larger contribution to the force, their effect is
of interest.  For these reasons, a direct test is necessary. Our
time-dependent Liouville equation can conveniently select a group of
states with a particular window.  Figure \ref{fig3}(a) includes states
from 5 ($3p^6$) to 100 (20 eV from the Fermi level), which covers the
core states, all the valence states and 66 conduction states. This is
the same data for our Fig. \ref{fig1}(d). By contrast,
Fig. \ref{fig3}(b) includes all the valence states but does not
include core states.  All the rest of the parameters, including the
laser parameters, are the same.  We see that their general trend
remains similar, but there is a clear difference: Figure \ref{fig3}(a)
has a negative $F_{1x}$, but Figure \ref{fig3}(b) has a positive
$F_{1x}$. The direction of $F_{1y}$ is also exchanged. This shows that
the core states do contribute.  {\clr We also understand why. Although
  the core states do not directly participate in the optical excitation,
  the force from the core state contains a correction term
  (Eq. \ref{core}) due to the effective potential $v_{eff}$. And this
  $v_{eff}$ depends on the electron density $\rho({\bf r})$. So during
  the laser excitation, once $\rho$ changes, the force due to the core
  state changes.}

Next, we investigate whether the spin-polarization scaling factor
\cite{jpcm16} affects the forces. This scaling is used in the
spin-polarized potential. For $\alpha$ electron, the potential is \be
V_\alpha(\br,t)=\frac{1}{2} \left [ (1+\xi(t))
  V_\alpha(\br)+(1-\xi(t)) V_\beta(\br) \right] \ee where
$\xi(t)=M(t)/M_0$ is computed from the spin moment $M(t)$. A similar
equation for $\beta$ electron. If we set $\xi(t)=1$, this means that
we do not have scaling.  If $\xi(t) <1$, the potential is less
spin-polarized.  Figure \ref{fig3}(c) has neither the core state nor
the spin-polarization scaling. We see that those rapid oscillations in
the force are suppressed. The force directions are now similar to
Fig. \ref{fig3}(a). Finally, we remove the spin-polarization rescaling
but keep the core states. Figure \ref{fig3}(d) shows that these rapid
oscillations return, and in addition the force is smaller. 

In summary, we should mention that our calculations are extremely time
consuming, 3-4 months to finish, since for each time step we
self-consistently converge the results. During the time when the laser
peaks, the convergence criteria are lowered because a large number of
conduction states are occupied and the self-consistency becomes
difficult. In traditional TDDFT, there is no self-consistency, but in
our time-dependent Liouville density functional theory, we have this
extra step.  There is another problem that is associated with the
Wien2k code. In the excited states, the electrons become very
delocalized, so we have to put some electrons as the background
charge. This background charge is not used for
Figs. \ref{fig3}(a)-(c), but is used for Fig. \ref{fig3}(d). For the
moment, we do not find a viable method to overcome this difficulty. A
planewave code may help, but it often uses the pseudopotential, where
the core states are replaced by a pseudopotential and the effect of
the core states on the force on atoms during laser excitation is not
possible to investigate.

\subsection{fcc nickel}

\begin{figure}
  \includegraphics[angle=0,width=1\columnwidth]{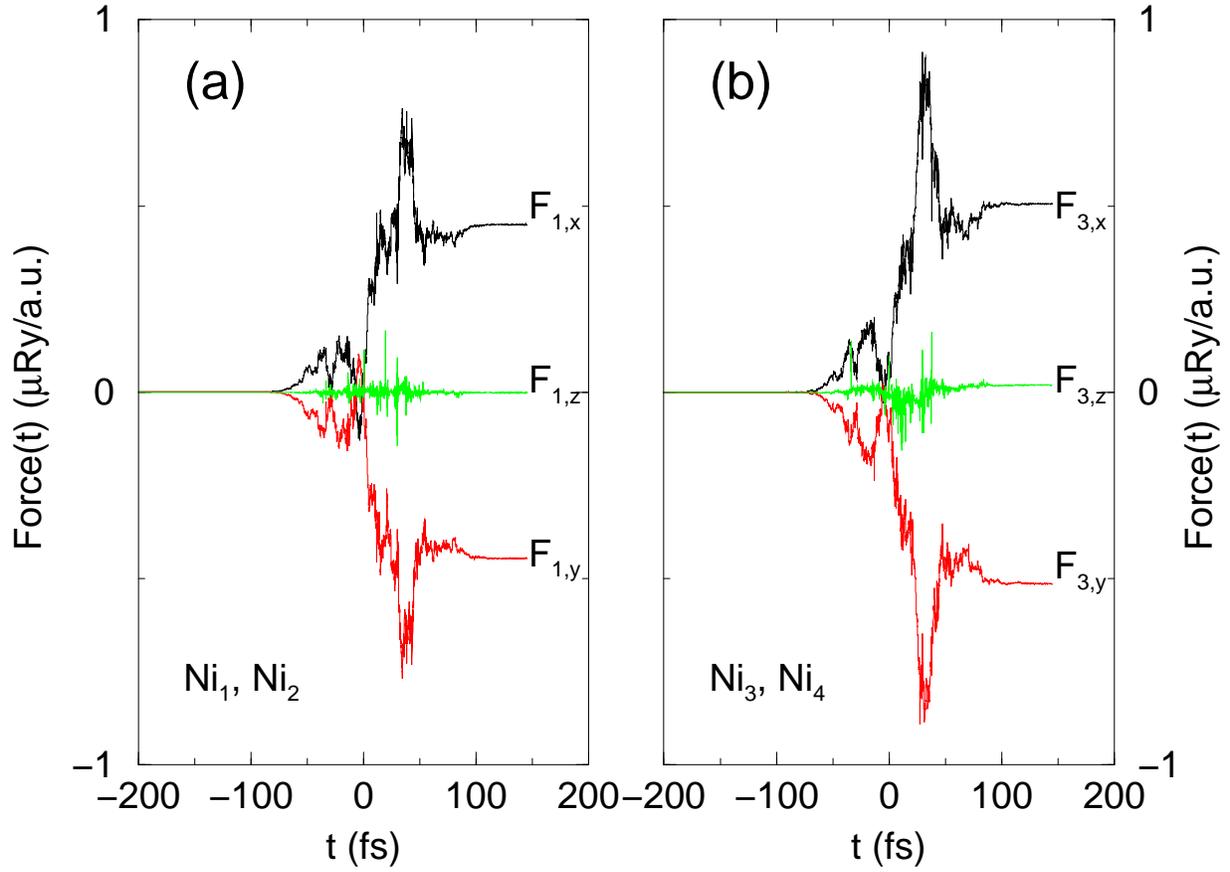}
  \caption{Forces in bulk fcc Ni calculated with a simple cubic cell
    and four atoms in total.  (a) Forces on Ni$_1$. The force on
    Ni$_2$ is similar and not shown.  To demonstrate the accuracy of
    our method, we remove all the symmetry operations, except the
    identity matrix.  The $k$ mesh is $14\times 14\times 14$.  $R_{\rm
      MT}K_{\rm max}$ is 7.  $\tau=60$ fs. $A_0=0.015 ~\rm
    Vfs/\AA$. $\hbar\omega=1.6$ eV. We use the circularly polarized
    light $\sigma_{xy}$. (b) Forces on Ni$_3$. The force on Ni$_4$ is
    similar and not shown.}
\label{fig2}
  \end{figure}

In order to demonstrate that our conclusion is not limited to the
Ni(001) monolayer, we also compute the force on bulk fcc nickel. For
the same reason as above, we adopt a simple cubic cell as a supercell
for fcc Ni, where we have four atoms per cell and remove all the
symmetry operations, except the identity matrix.  Figure \ref{fig2}(a)
shows the forces on atoms Ni$_1$ and Ni$_2$. Since forces on Ni$_1$ and
Ni$_2$ are similar, we only show that on Ni$_1$. They do not gain an
appreciable value until -70 fs.  $F_{1,x}$ points in the $x$ axis,
while $F_{1,y}$ points in the $-y$ axis. The magnitude of $F_{1,z}$ is
much smaller. At 0 fs when the laser peaks, the force drops to
zero. The force peaks around 33 fs, after which it stabilizes around
0.5 $\mu$Ry/bohr.  Figure \ref{fig2}(b) shows the forces on Ni$_3$ and
Ni$_4$. The general feature is similar.  We can estimate how far atoms
move, provided the force remains constant. Take the force at 0.5
$\mu$Ry/bohr as an example, and we obtain the acceleration $a$ is
$0.2114\times 10^{12}$ m/s$^2$. For 1 ps, the atom moves 0.1057 pm,
far less than 1.9 pm postulated by Tauchert \et
\cite{tauchert2022}. If we used their time scale of 125 fs (8 THz),
the displacement would fall below the thermal
disturbance. Importantly, our theory predicts that $F_{1,x}$ and
$F_{1,y}$ are 180$^\circ$ out of phase, with the resultant force along
the $[1,\bar{1},0]$ direction.  These forces are unlikely to lead to a
circular motion of Ni atoms around the lattice point, because they are
not centripetal in any sense. Instead, a regular vibration is
expected. This raises a question whether their polarized phonon really
exists.

\section{Conclusions}

In contrast to all the prior studies, we have directly computed the
laser-induced forces on atoms during ultrafast demagnetization. We
find that the force arises around -50 fs before the laser pulse peaks
at 0 fs. Due to the strong charge fluctuation, the initial forces are
highly oscillatory, and peak around 50 fs, settling down after 100
fs. Although we have observed that the force does change its
direction, it does not appear to be centripetal, as would be expected
from the polarized phonon picture \cite{tauchert2022}. Quantitatively,
we find the force magnitude is at least one order of magnitude weaker
than the early estimate \cite{tauchert2022}. This conclusion is true
for both the Ni(001) monolayer and fcc nickel, within the limit of our
current theory.  Interestingly, a recent study in He atoms reported
the long-lasting orbital moment of the electron \cite{watzel2022}, not
phonon. Could the polarized phonon turn out to be the electron
contribution, not the lattice contribution? This likelihood scenario
is supported by our theory and also the fact that the experiment
employed the electron diffraction, but it can only be resolved with
further experimental and theoretical investigations
{\clr \cite{jpcm22,mankovsky2022}.}

\acknowledgments

We would like to thank the helpful communication with Dr. Peter Baum
(Konstanz, Germany) on their paper \cite{tauchert2022}.  GPZ and YHB
were supported by the U.S. Department of Energy under Contract No.
DE-FG02-06ER46304.  Part of the work was done on Indiana State
University's quantum and obsidian clusters.  The research used
resources of the National Energy Research Scientific Computing Center,
which is supported by the Office of Science of the U.S. Department of
Energy under Contract No. DE-AC02-05CH11231.

$^*$guo-ping.zhang@outlook.com
https://orcid.org/0000-0002-1792-2701

\end{document}